\documentclass[aps,prl,showpacs,twocolumn]{revtex4-1}
\usepackage{graphicx}
\usepackage{textcomp}
\usepackage{txfonts}
\usepackage[dvipsnames]{xcolor}
\usepackage{epstopdf}

\begin{document}

\title{Dynamic Displacement Disorder of Cubic BaTiO$_3$}

\author{M. Pa\'{s}ciak$^{\rm a}$}
\email{pasciak@fzu.cz}
\author{T. R. Welberry$^{\rm b}$}
\author{J. Kulda$^{\rm c}$}
\author{S. Leoni$^{\rm d}$}
\author{J. Hlinka$^{\rm a}$}
\email{hlinka@fzu.cz}
\affiliation{
$^{\rm a}$
Institute of Physics of the Czech Academy of Sciences\\%
Na Slovance 2, 182 21 Prague 8, Czech Republic\\
$^{\rm b}$Research School of Chemistry, Australian National
  University, Canberra ACT 0200, Australia\\
$^{\rm c}$Institut Laue-Langevin, BP 156, 38042 Grenoble Cedex 9, France\\
$^{\rm d}$Cardiff University, School of Chemistry, Park Place, CF10 3AT, Cardiff, UK\\
}


\begin{abstract}
The three dimensional distribution of the X-ray diffuse scattering intensity of BaTiO$_3$ 
has been recorded in a synchrotron experiment and simultaneously computed using molecular dynamics simulations of a shell-model. Together these have allowed the details of the disorder in paraelectric BaTiO$_3$ to be clarified. The narrow sheets of diffuse scattering, related to the famous anisotropic longitudinal correlations of Ti ions, are shown to be caused entirely by the overdamped anharmonic soft phonon branch. This finding demonstrates that the occurrence of narrow sheets of diffuse scattering agrees with a displacive picture of the cubic phase of this textbook ferroelectric material.
\end{abstract}


\maketitle

The narrow lines of diffuse scattering, discovered half a century ago in X-ray photographs of BaTiO$_3$ crystal, 
have remained a puzzle for several generations of physicists interested in the nature of ferroelectricity ~\cite{Comes1968,Comes1970,Huller1969,Harada1967,Harada1965,Takesue1995,Ravel1998,Hlinka2008,Ravy2007,Liu2007,Pasciak2010a,Senn2016}. 
It was
soon understood that the observed diffuse scattering 
reflects a very peculiar nanoscale displacement disorder of Ti ions~\cite{Comes1968,Comes1970,Harada1967}.

\begin{figure}[!tbp]
\includegraphics[width=62mm]{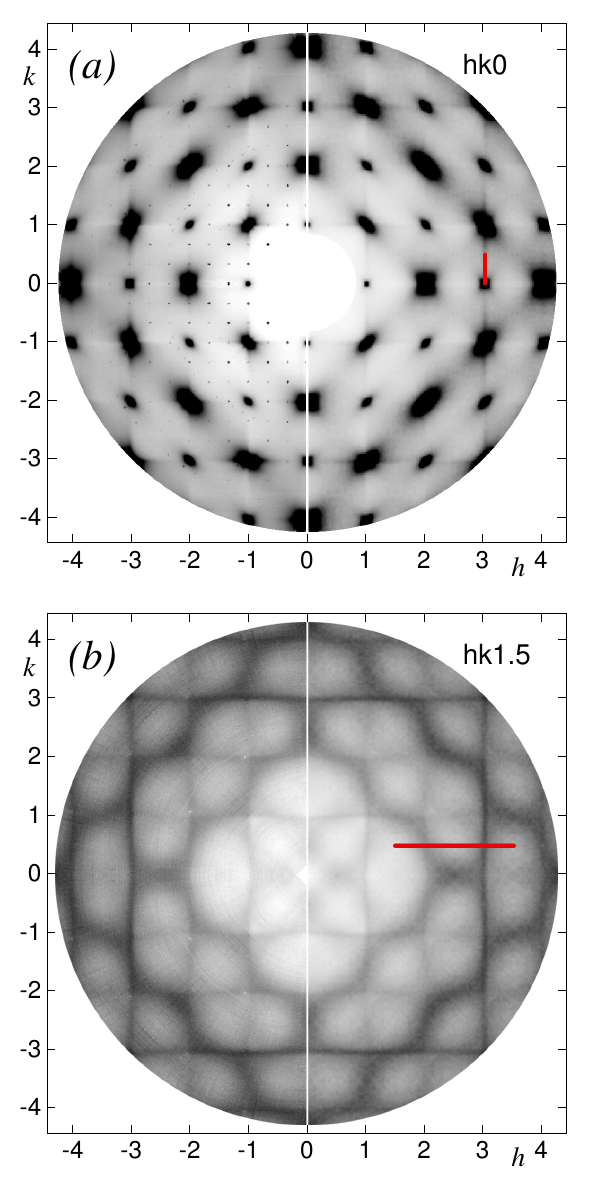}
 \caption{Diffuse scattering of cubic BaTiO$_3$ single crystal at T$_{\rm C}$ + 100~K within (a) $hk0$, (b)  $hk1.5$  planes of the momentum space. Left semicircles show experimental data, right ones are calculated (see the text). Darker contrast means higher  intensity. 
 The vertical and horizontal diffuse lines are caused by the  short range Ti-ion displacement correlations. 
  Two highlighted segments relate to the paths explored in Figs.\,3 and 4. Sharp tiny spots in non-integer positions in left semicircle of (a) come from ${\lambda}/{3}$ contamination of the beam.
  } \label{fig1}
\end{figure}

An example of such diffuse scattering is the $k=3$ dark horizontal straight line in an 
X-ray diffuse scattering image shown in Fig.\,1a. 
The analysis of similar images obtained for BaTiO$_3$ and related perovskites indicated that Ti ion displacements parallel 
to a given Ti-O-Ti bond chain are correlated along this chain up to distance of the order of 5-10\,nm, while there is little correlation 
between the  displacements perpendicular to the chain~\cite{Comes1968}.
Since then 
until the present these correlated displacements are considered as the key ingredient in the phase transition of BaTiO$_3$ and they are 
deemed responsible  for a range of nonstandard phenomena occurring  even in the paraelectric cubic phase \cite{Tai2004,Pugachev2012,Ko2011,Senn2016,Dulkin2010,Malinovsky2013}.

Such correlated displacements, hereafter 
referred to as chain correlations \cite{Comes1968}, can be well explained 
if it is assumed
that each Ti cation is at any moment off-centered with respect to the surrounding oxygen octahedral 
cage, and displaced towards one of its eight facets \cite{Comes1968}. The 
8-site off-center model gives a tractable framework for many quantitative 
considerations \cite{Comes1968,Kopecky2012,Zalar2003,Takahashi,Pirc2004,Zalar2005}
and, in particular, it allows 
an explanation to be given as to why these chain correlations eventually vanish in the rhombohedral ground state phase \cite{Comes1968,Pasciak2010a,Senn2016}.

Nevertheless, 
it has been argued that similar chain correlations might be induced simply by the low-frequency phonon modes \cite{Huller1969,Harada1965,Yu1995,Takesue1995,Harada1967,Harada1971,Zhang2006}.
Since the available conventional X-ray scattering data do not provide any direct information about the timescale of the chain correlations,
many other techniques were employed to try to solve this controversy. 
These efforts yielded conflicting conclusions. On the one hand, many experiments, in particular the local probe methods,  supported the pronounced 8-site off-centering \cite{Pirc2004,Ravel1998,Ravy2007},
while the spectroscopic methods designed to probe collective excitations typically indicated 
that the classical soft mode picture of the phase transition is more 
appropriate \cite{Inoue,Vogt1982,Harada1971,Presting1983}.

The most convincing spectroscopic evidence for order-disorder polarization dynamics has been  found in the THz-range frequency response of the dielectric permittivity  of the {\it tetragonal} ferroelectric phase of BaTiO$_3$ \cite{Hlinka2008}. Its spectrum  shows an additional relaxational polar mode,  well separated from the three normal IR active phonon modes expected in the cubic perovskite crystal. In addition, theoretical modeling indicated that the characteristic frequency of this extra relaxational polar mode roughly matches the rate of the single-ion 
hopping between the inequivalent off-centered Ti positions (favored and disfavored by the spontaneous polarization, respectively) \cite{Hlinka2008}.

In contrast, there is no similarly obvious spectroscopic evidence for the inter-site 
hopping
dynamics in the {\it cubic} phase of BaTiO$_3$. 
While detailed fitting of the paraelectric soft mode spectral response with a single damped harmonic oscillator is not fully satisfactory,  there is certainly no well-separated central peak in the paraelectric spectra that could be ascribed straightforwardly to the dynamics of the inter-site jumps among the eight off-centered Ti positions \cite{Ponomareva2008, Weerasinghe2013, Inoue}.
Moreover, it 
has not yet been established
how the  spectroscopic results relate to the diffuse scattering observations:
are the THz-range polar excitations responsible  also for the X-ray diffuse scattering planar sheets?

In order to clarify the phase transition mechanism of this  textbook ferroelectric substance in a broadly accessible manner, we have repeated the original X-ray diffuse scattering experiment with currently available means. The measurement was carried out at 500\,K, {\it i.e.} about 100\,K above T$_{\rm C}$, where most of the recently reported peculiar properties \cite{Pugachev2012,Roth2017,Aktas2013,Zhang2015,Tsuda2016},
often forbidden by the cubic symmetry,   should be either absent or negligible. 
A combination of 
the currently available synchrotron source experimental data with the contemporary molecular dynamics modeling yields a very clear-cut picture about the dynamics of the chain correlations in the cubic phase of BaTiO$_3$. In particular, while the planar sheets of X-ray diffuse scattering are shown to be caused by the soft phonon branch scattering,  the signatures of a strongly anharmonic local potential with 8 most probable off-centered sites is clearly present in the material as well.

The experimental data were obtained
at the Advanced Photon Source 11ID-B beamline using Perkin-Elmar amorphous silicon detector and incident X-ray beam with the energy of 58 keV (0.2127~\AA{}). During the measurement, the BaTiO$_3$ single crystal sample, held at the temperature of 500\,K, was rotated with
$\omega = 0.25^{\circ}$ step in a way allowing for a systematic coverage of
of the reciprocal space. The reconstruction of the reciprocal space planes was done with the program \textsc{Xcavate}~\cite{Estermann1998}, after indexing the data within the cubic $Pm\overline{3}m$ spacegroup with the lattice parameter $a$=4.01~\AA{} and after removal of the known artifacts~ \cite{Welberry2005,Welberry2004}. To the best of our knowledge, such 
detailed quantitative information about the diffuse scattering in cubic phase of BaTiO$_3$ has not been reported yet.

Theoretical diffuse scattering intensity maps were derived from computer simulations yielding temporal evolution of atomic positions, contained in a box comprising 50$^3$ BaTiO$_3$ unit cells. The trajectory was obtained from
 molecular dynamics simulations conducted with the \textsc{dlpoly} software~\cite{Todorov2006}, using 
 an \textit{ab-initio}-based shell-model interatomic potential  taken from 
 Sepliarsky~\textit{et al.}~\cite{Sepliarsky2005}. 
The timestep was 0.2~fs.  After the appropriate equilibration~\cite{nst_nvt} the production trajectory
was obtained from about 100\,ps of NVE (constant volume-constant energy) ensemble run.
The ferroelectric phase-transition temperature in the model 
occurs 
at $T_{\rm C, theor}\sim 300$\,K;  for 
a reasonable comparison with the experiment, the calculations were conducted at $T=T_{\rm C, theor}+100$\,K. Theoretical diffuse scattering intensities 
were obtained with the \textsc{discus} program~\cite{Proffen1997}.
Intensity maps 
presented in 
this work 
were determined as an average of 48 images 
each calculated as a cubic average from one 
snapshot of the structure trajectory. 
In order to directly identify the dynamics of the scattering processes involved, we have used the information stored in the  MD trajectory and numerically evaluated the scattering  efficiency at a given  momentum transfer $\hbar {\bf  Q}$ according to the associated energy transfer  $\hbar \omega$, in terms of the $S({\bf Q},\omega)$ scattering function with 
0.4~meV energy resolution at several points and paths in the momentum space,  using the nMoldyn program~\cite{Hinsen2012}. 

The comparison of the experimental and theoretical results for diffuse scattering intensity distributed 
within the $hk0$ and $hk1.5$ reciprocal space planes of cubic BaTiO$_3$ is displayed in Fig.~\ref{fig1}. Left semicircles of each intensity map in Fig.~\ref{fig1} show  experimental results, 
right semicircles 
show molecular dynamics prediction.

The integer-coordinate points in the $hk0$ reciprocal space plane correspond to the 
vertices of the reciprocal lattice of the cubic perovskite structure.
The dark diffuse spots located at these reciprocal lattice points are due to the usual thermal diffuse scattering by the thermally-activated 
low-frequency 
acoustic phonon modes. Each of these spots 
appears as either  an individual elongated ellipsoid or as a pair of differently oriented overlying ellipsoids 
(evoking a butterfly shape). The shape and the orientation of these ellipsoids can be well understood from the anisotropy of the elastic tensor of the cubic BaTiO$_3$ crystal \cite{Harada1967,Harada1971}.
  
There is also a somewhat weaker diffuse scattering intensity  emanating from the long axes of the ellipsoidal spots
that seems to mutually connect the diffuse spots 
into continuous diffuse scattering stripes (for example, the $h+k=4$  intensity stripe in Fig.~\ref{fig1}a).
More precisely, this weak diffuse scattering intensity 
is concentrated around all reciprocal space planes 
coincident with the facets of reciprocal space octahedra, defined by $|h|+|k|+|l| \leq 2n$, where $n=1,2,3$, etc..
For example, the diffuse scattering 
localized near
the $h+k+l=4$ plane appears as the dark stripe rippled around the  $h+k=2.5$ reciprocal line in Fig.~\ref{fig1}b, and as the $h+k=4$ dark stripe in Fig.~\ref{fig1}a.
It is natural to ascribe 
these diffuse stripes to the low-frequency phonon modes
as well.
In fact, the stripe intensity  roughly scales with the intensity of 
the adjacent thermal diffuse scattering spots, and the latter follows the known structure factor variation typical for acoustic modes. It has been thus proposed that the stripes and spots are {\it together} forming an "acoustic component" of the diffuse scattering $S_{\rm A}$ \cite{Harada1967}. 

\begin{figure}
\includegraphics[width=80mm]{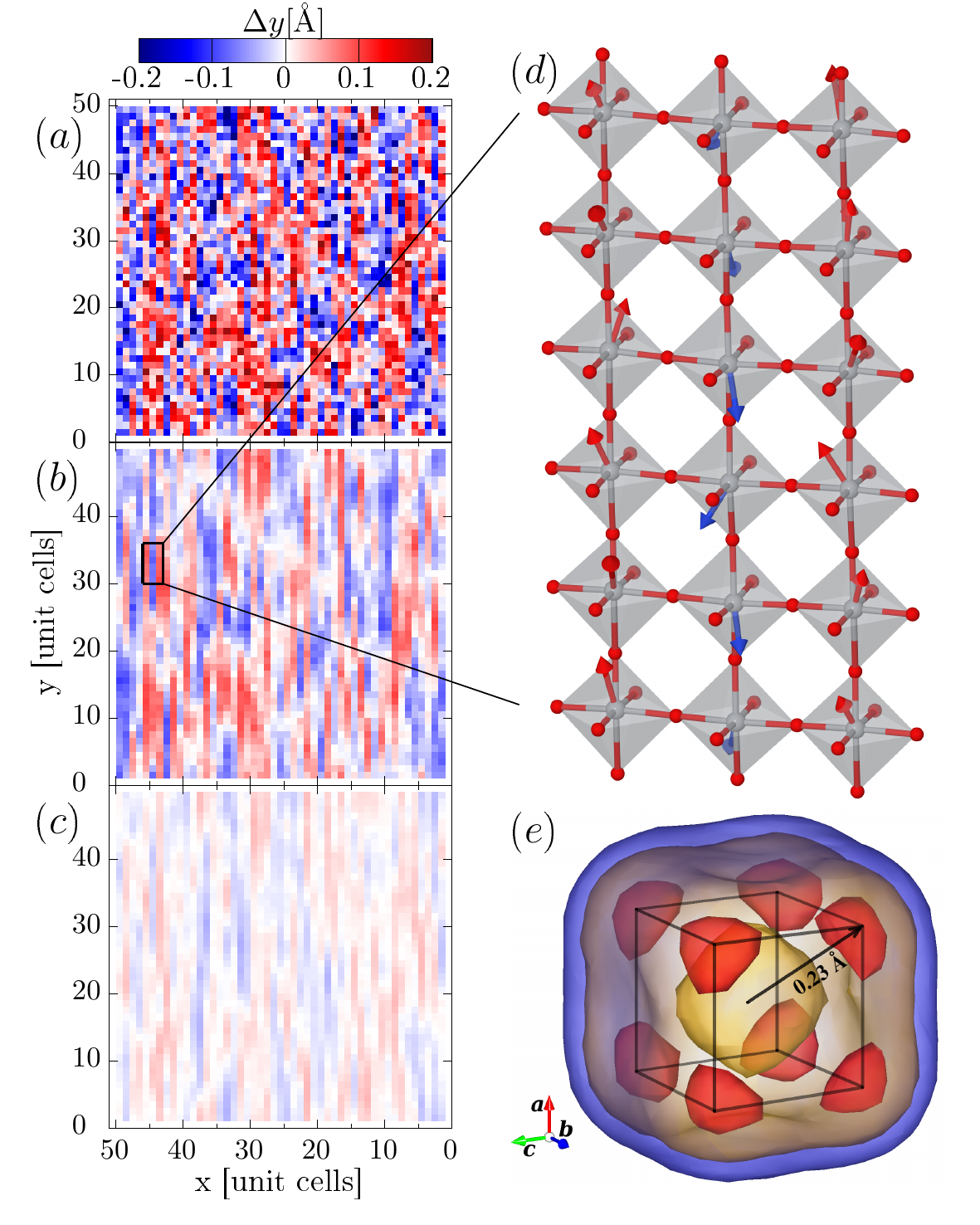}
 \caption{Distribution of the $y$-components of the Ti ion displacement within a selected $xy$ layer of the simulated crystal box: (a) instantaneous displacements,  (b) displacements averaged over 500\,fs period, (c) displacements averaged over 4\,ps period. Panel (d) shows a fraction of the simulated atomic structure with arrows showing magnified vector of the local Ti displacement. (e) Isosurface plot of probability distribution for Ti atom position with respect to the center of the oxygen octahedron. The 8 most likely positions are marked as corners of the superposed cube with about 0.3\,\AA~long edges. } \label{fig2}
\end{figure}

In contrast, the scattering related to the Ti chain correlations is  restricted to the 
closest 
vicinity of the $|h|=n$, $|k|=n$ and $|k|=n$ ($n=1,2,3...$)  reciprocal planes. 
Consequently,  it appears just as a set of extremely narrow intensity lines, parallel to the Cartesian axes in the images of Fig.~\ref{fig1} (the dark horizontal line at $k=3$, for example). Visibly, the intensity of these lines does not 
scale with the intensity of the neighboring acoustic diffuse spots. Rather, the observed intensity variations testify that these diffuse lines stem from the disordered optic-like displacements, representing the relative
 displacement of Ti ion with respect to the neighboring O and Ba ions \cite{Harada1967}. The debated question is whether this optical part of the diffuse scattering, $S_{\rm{O}}$,  simply reflects the dispersion of low-frequency soft optic modes, or whether this diffuse scattering arises 
 due to some extra source of  Ti chain correlations, which is 
 inherently linked to the 8-site Ti off-centering.
 
The calculated trajectory allows to inspect directly the Ti ion displacements with respect 
the
oxygen octahedral cages. The 3-dimensional histogram of Ti ion positions with respect to the center of its oxygen octahedron shows a very smeared probability distribution, similar to that found in Ref.\,\cite{Qi2016}, nevertheless with a shallow local minimum in the center and eight maxima at the anticipated~\cite{Polinger2013,Ravy2007,Senn2016,Pirc2004} off-center positions (see Fig.\,2e). 
Still, only about 5 percent of the Ti ions actually fall within these probability peaks, most of the Ti ions are distributed around them.

An example of the instantaneous distribution  of the relative Ti-O$_6$ displacements in the real space is shown in Fig.\,2a (only the $y$-component is displayed). Anisotropic chain correlations between neighboring sites  are better revealed in the picture showing  displacements 
averaged over the time of 500\,fs (Fig.\,2b). 
For example, the $y$-component of a particular Ti local displacement tends to be parallel to that of its neighbors in the $y$-direction,
but there is little correlation in the perpendicular directions~\cite{Pasciak2010a,Geneste2011}. Equivalent correlations obviously hold for displacements along the $x$ and $z$-axes.
Averaging over a
longer period leads to a similar picture but with a significant reduction of the displacement magnitudes  (see the 4\,ps averaging in Fig.\,2c). This indicates the picosecond lifetime of these correlations.

\begin{figure}
 \includegraphics[width=80mm]{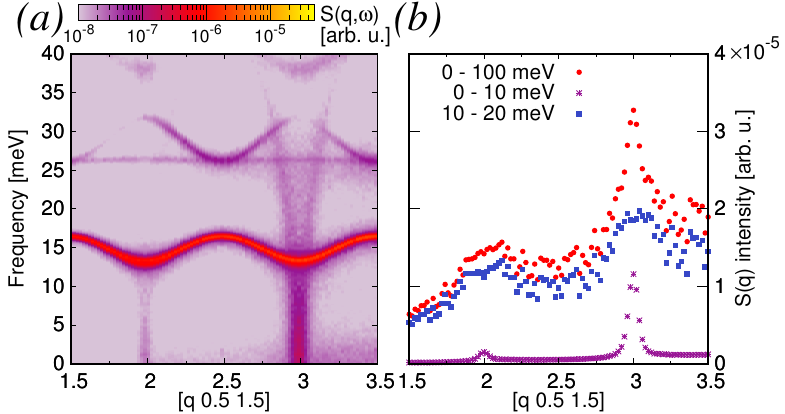}
 \caption{Dynamical analysis of the diffuse scattering intensity:  (a) Map of $S(q,\omega)$ in the [$q$ 0.5 1.5] direction,
  (b) 
  X-ray scattering intensity determined from integration of $S(q,\omega)$ over the indicated frequency intervals.
  } \label{fig4}
\end{figure}

The dynamical nature of the thermal excitations involved in these Ti correlations can be more quantitatively understood from the frequency dependence of the $S({\bf Q},\omega)$ scattering function. The profile of $S({\bf Q},\omega)$  calculated for the [q 0.5 1.5] trajectory in the momentum space 
is shown in Fig.\,3a. It passes through two 
$S_{\rm A}$ stripes located 
around $q=2$ and $q=3$, 
as well as through the  $S_{\rm O}$ diffuse line at $q=2$ and $q=3$ (see the segment marked in Fig.\,1b).
At the simulated temperature, phonons with frequencies above 40\,meV are barely populated so that
the overall scattering is given by integration of $S({\bf Q},\omega)$ over the displayed frequency range. Fig.\,3b shows the full integral as well as the contributions obtained by integration with 0-10 and 10-20\,meV. 
Two key conclusions can be drawn here. First,  the $S_{\rm A}$  diffuse scattering stripes are mostly caused by the dispersion of the rather flat phonon branches in the 10-20\,meV frequency region (it is barely seen in Fig.\,3a, but the modes with lower frequencies are more populated and thus contribute more to the integral intensity). 
Second, the sharp peak at $q=3$ in Fig.\,3b arises due to the strongly dispersive phonon branch, apparently dropping from more than 40~meV down to a zero
frequency (at $q=3$, see Fig.\,3a). Thus, it is this branch that is behind  $S_{\rm O}$ and the discussed Ti-chain displacement correlations. 

Finally, it is interesting to inspect the $S({\bf Q},\omega)$
scattering function  {\it along} the sheet of diffuse scattering.
The $S({\bf Q},\omega)$ calculated for the  [$q$ 3 0] 
path
(Fig.\,4a) clearly reveals dispersion of 
both 
the transverse acoustic branch and an optic branch in the 22-25\,meV frequency region.
On the top of it, there is a marked  scattering component distributed around the zero energy transfer channel ($\omega=0$), indicating  an overdamped or a relaxational mode. 

In order to eliminate the scattering by 
overlapping low-frequency acoustic mode from the spectra,
we have also calculated an auxiliary scattering function $A({\bf Q},\omega) =S({\bf Q},\omega)_{Ba}+S({\bf Q},\omega)_{BaTi}+S({\bf Q},\omega)_{Ti} $, where $S_{Ba}$, $S_{Ti}$ and $S_{BaTi}$ 
are scattering functions calculated 
considering the scattering by Ba ions only, by Ti ions only or by Ba and Ti ions only, respectively. This auxiliary scattering function suppresses the contribution of the long-wavelength acoustic branch (Fig.\,4b) so that the spectral shapes of the overdamped optic-like excitation  can be then followed  even in the closest vicinity of the Brillouin zone center (Fig.\,4c). In particular, the half width at half maximum (HWHM) of the $q=0.02$ spectrum is
about $1-2$\,meV, which implies that the imaginary susceptibility has a maximum around this frequency \cite{AlZein}. 
This value corresponds to the position of the soft mode-related dielectric loss function maximum of BaTiO$_3$ (at $T_{\rm C}+100$\,K), as determined by optical spectroscopy methods  \cite{Vogt1982,Ponomareva2008,Weerasinghe2013}.
 In other words, this analysis confirms that the leading contribution to the X-ray diffuse scattering by the Ti-chain correlations coincides with the scattering by the very same THz-range soft-phonon branch, that determines the high dielectric permittivity of BaTiO$_3$.

\begin{figure}
 \includegraphics[width=80mm]{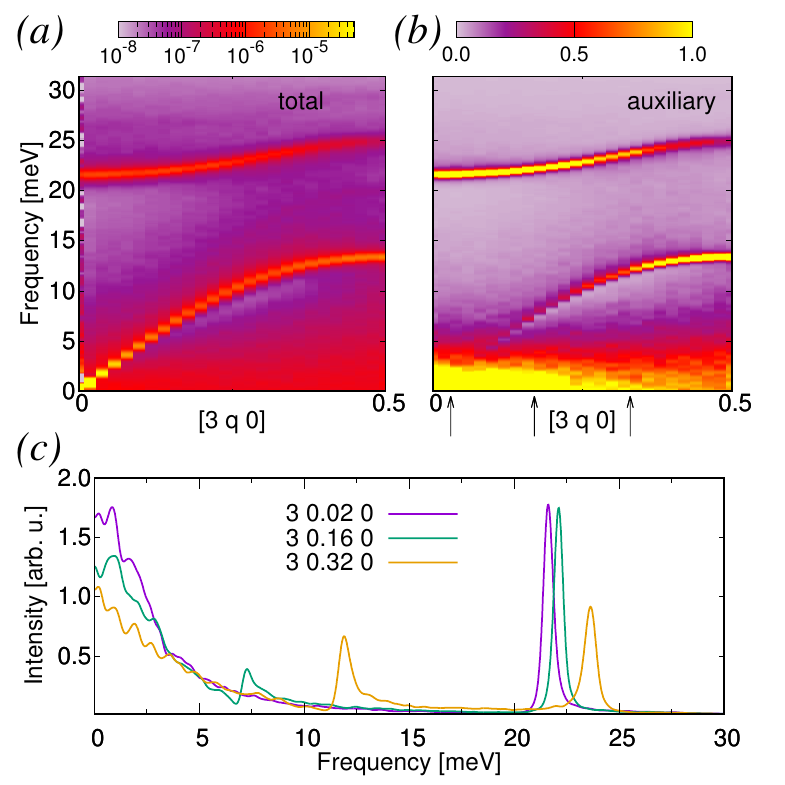}
 \caption{Dispersion maps in the [3 $q$ 0] direction: (a) total, displayed using log scale, (b) linear combination of partial $S(q,\omega)$: $S_{Ba}+S_{BaTi}+S_{Ti}$ that unveils optic part of the signal. (c) Optic spectra at different $q$ values indicated by arrows on the map in (b).} \label{fig4}
\end{figure}

In summary, this Letter reports a synchrotron X-ray measurements of diffuse scattering in cubic phase of BaTiO$_3$ with a considerably enhanced momentum space coverage and dynamical contrast. Large-scale molecular dynamics simulations allowed us to calculate maps of diffuse scattering in excellent 
agreement with those observed experimentally. The calculated high-temperature trajectory allowed us to access the dynamical profiles of the observed reciprocal space features.

The obtained results agree well with a range of previous experiments and calculations, for example with the inelastic neutron scattering studies \cite{Harada1971} and computer simulations~\cite{Qi2016,Noordhoek2013,Geneste2011,Hlinka2008,Ponomareva2008,Tinte2000}. Full discussion of the results is beyond the scope of this letter and will be presented elsewhere.
On the other hand, 
we hope that the present clarification of the order-parameter dynamics in BaTiO$_3$ can be of help in various considerations about the polar nanoregions, about peculiarities of the precursor phenomena etc. in perovskite ferroelectrics in general.

Let us stress that the presented methodology allows one to go beyond the harmonic approximation in 
the 
analysis of phonons and phonon-related scattering. This gives us a possibility to clarify several challenging aspects of the phase transition in BaTiO$_3$.
We not only show  that the sufficiently soft phonon modes  
are present there, that their $\Gamma$-point spectrum matches the known macroscopic dielectric spectrum and that the associated dispersion in the momentum space is steep enough to explain the observed sharpness of the diffuse scattering lines, but we also demonstrate that these anharmonic phonons are really the cause of  these diffuse scattering lines, in the sense that there is no space for any other,  comparably intense contribution, that would need to be added to build up the resulting X-ray diffuse scattering intensity of these lines. 

\begin{acknowledgments}
 This work was supported by the Czech Science Foundation (project no. 13-15110S). Use of
the Advanced Photon Source was supported by the U. S. Department
of Energy, Office of Science, Office of Basic Energy
Sciences, under Contract No. DE-AC02-06CH11357.
The computational part of this research was undertaken on the 
NCI National Facility in Canberra, Australia, which is supported by the Australian Commonwealth Government. We would like to thank Dr Aidan Heerdegen for his help during the experiment.

\end{acknowledgments}


\begin{thebibliography}{14}
\bibitem{nst_nvt} The system 
was
consecutively equilibrated 
over NST (constant stress-constant temperature) and NVT (constant volume-constant temperature) ensembles 
for stabilization of volume and temperature, respectively. As a slow drift of shell temperature has been observed, the shells have been allowed to reach the equilibration temperature.
\end{thebibliography}
\end{document}